\documentclass[prd,aps,preprintnumbers,fleqn,showpacs,nofootinbib,superscriptaddress]{revtex4}

\usepackage[utf8]{inputenc}
\usepackage[T1]{fontenc}
\usepackage{slashed}
\usepackage{times}

\usepackage{amssymb,amsfonts,amsmath,amsthm}
\usepackage{dsfont,bbm}
\usepackage{dcolumn}

\usepackage{graphicx}
\usepackage[caption=false]{subfig}

\begin{document}

\title{Color entanglement effect in collinear twist-3 factorization}

\author{Jian~Zhou}
\affiliation{School of Physics and Key Laboratory of Particle Physics and Particle Irradiation
(MOE),  Shandong University, Jinan, Shandong 250100, China}

\begin{abstract}
\noindent We study color entanglement effect for T-odd cases in collinear twist-3 factorization. As
an example, we compute the transverse single spin asymmetry for direct photon production in pp
collisions in  pure collinear twist-3 approach. By analyzing the gauge link structure of the
collinear gluon distribution on unpolarized target side, we demonstrate how color entanglement
effect arises in the presence of the additional gluon attachment from polarized projectile. The
result is consistent  with that obtained from a hybrid approach calculation.
\end{abstract}

 \maketitle
\section{introduction}
One of the key steps involved in proving QCD factorization theorems is to decouple longitudinal
gluon exchange between  active partons and  the remanets of projectile/target nucleons in a high
energy scattering process. By invoking the Ward identity argument, longitudinal gluon exchange to
all orders can be absorbed into gauge link that ensures gauge invariance of the operator
definitions of parton distribution functions. In the context of transverse momentum dependent(TMD)
factorization framework~\cite{Collins:1981uk}, depending on color flow in a hard scattering, TMD
parton distributions could possess quite complicated gauge link
structure~\cite{Bomhof:2004aw,Bacchetta:2005rm,Bomhof:2006dp,Collins:2007nk}. A generalized TMD
factorization  was proposed~\cite{Bacchetta:2005rm,Bomhof:2006dp} to express cross sections  in
terms of process dependent TMDs when computing physical observables that are sensitive to incoming
parton transverse momenta.

However, a further investigation~\cite{Rogers:2010dm} reveals that it is not possible to
disentangle simultaneous longitudinal gluon attachments from both nucleon sides in a
nucleon-nucleon collisions provided that color flow is nontrivial in final state. This phenomena,
commonly refereed to as color entanglement, originates from the non-Abelian  feature of QCD as a
gauge theory. It prevents us from describing parton transverse momentum with separate correlation
functions for each external nucleon, and thus leads to the breakdown of generalized TMD
factorization. The phenomenological implications of color entanglement effect have been explored
from both theoretical and experimental
sides~\cite{Buffing:2011mj,Buffing:2012sz,Rogers:2013zha,Adare:2016bug}.

On the other hand, it is not yet entirely clear whether color entanglement effect shows up in
nucleon-nucleon collisions in collinear factorization calculation. Though it is usually believed to
be absent in  leading twist collinear factorization, at twist-3 level, one has to take into account
an additional gluon re-scattering which makes color flow more complicate, and could potentially
give rise to color entanglement effect. In fact, the transverse single spin asymmetry(SSA) for
prompt photon production in pp collisions computed in genuine collinear twist-3
factorization~\cite{Qiu:1991wg,Kouvaris:2006zy} differs from that obtained in the hybrid
approach~\cite{Schafer:2014zea} by terms proportional to a new gluon distribution $G_4$, whose
emergence can be attributed to color entanglement effect. To some extend, the hybrid approach is a
more complete method in the sense that gluon re-scattering on the unpolarized target side has been
summed to all orders. One thus has good reason to believe that the color entanglement effect
related to the $G_4$ term contribution  are missing in the conventional collinear twist-3
calculation.

The objective of this paper is to explicitly work out color structure for the diagrams with
simultaneous longitudinal gluon attachments from both incoming nucleon sides within the pure
collinear twist-3 factorization framework.
 To the best of our knowledge,  the gauge link structure of leading
twist collinear parton distributions on unpolarized target side has never been carefully examined
in the presence of additional gluon attachment from polarized projectile.  To identify color
entanglement effect and compare with the hybrid approach, it is sufficient to take into account one
 longitudinal gluon attachment from unpolarized target on each side of cut,
 since this is the lowest nontrivial order where $G_4$ receives nonvanishing contribution. As
 discussed
 in Sec. III, the gluon distribution $G_4$ indeed enters into the spin dependent cross section in
 pure collinear twist-3 approach when going beyond one gluon exchange approximation. As expected,
 we verified  that the hybrid approach and the collinear twist-3 approach yield the same result in the
 collinear limit at the order under consideration.

 The paper is structured as follows. In the next section, we briefly review  the conventional
 collinear twist-3 calculations for the SSA in direct photon production and the derivation of gauge
 link of the collinear gluon distribution in leading twist collinear factorization.
 In sec. III, we present our analysis for the gauge link  structure at twist-3 level in details,
 and recover the hybrid approach result.  In the end, we comment on more general cases and discuss
 possible extensions of the current work in sec. IV.

\section{Brief review of conventional calculations}
\begin{figure}[hbtp]
\begin{center}
\includegraphics[angle=0,scale=0.6]{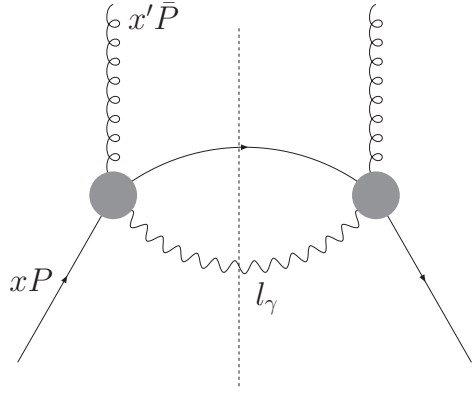}
\caption{Diagram contributing to direct photon production in pp collisions. Grey circles indicate
all possible photon line attachments.} \label{1}
\end{center}
\end{figure}
The dominant production mechanism for prompt photons in high energy collisions is Compton
scattering $gq \rightarrow\gamma q$ as shown in Fig.1. We start by introducing the relevant
kinematical variables and  assign 4-momenta to the particles according to
\begin{eqnarray}
 g(x' \bar P )\ + \ q(xP)\longrightarrow \gamma(l_\gamma) \  + \ q(l_q)
\end{eqnarray}
where $\bar P^\mu=\bar P^- n^\mu$ and  $ P^\mu= P^+ p^\mu$ with $n^\mu$ and $p^\mu$ being the
commonly defined light cone vectors, normalized according to $p \cdot n=1$.
 The corresponding unpolarized Born cross section reads,
\begin{eqnarray}
\frac{d^3  \sigma}{d^2 l_{\gamma\perp} dz }=
 \frac{\alpha_s \alpha_{em} }{N_c} \
\frac{z[1+(1-z)^2]}{l_{\gamma\perp}^4} \sum_q e_q^2 \int^1_{x_{min}} dx \ f_q(x)  x'G(x')
\label{colunp}
\end{eqnarray}
where $z\equiv l_\gamma \cdot n/(x P \cdot n) $ is the fraction of the incoming quark momentum $xP$
carried by the outgoing photon,
 and $l_{\gamma\perp}$ is the photon transverse momentum.
The meaning of the other coefficients should be self-evident. Note that $x'=\frac{xP \cdot l_q}{xP
\cdot \bar P-P \cdot l_\gamma}$ is a function of $x$;
 and $x_{min}$ is given by $x_{min}=\frac{P \cdot l_\gamma}{P \cdot \bar P-P \cdot l_q}$.
In the above formula, $f_q(x)$ and $G(x')$ are the usual integrated quark and gluon distributions,
respectively.

The operator definition of the collinear gluon distribution is given by~\cite{Collins:1981uk}
\begin{eqnarray}
 x' G( x')=\int \frac{d\xi ^{+}}{2\pi \bar P^{-}}e^{-ix' \bar P^{-}\xi^{+}} \langle
P|F^{-\mu}_a(\xi^{+} )\mathcal{\tilde L}_{ac} F^{-\mu}_c(0)|P\rangle \ , \label{g1}
\end{eqnarray}
where $F^{-\mu }_a$ is the gauge field strength tensor and $ \mathcal{\tilde
L}_{ac}=\mathcal{P}\exp\{-g\int_{0}^{\xi ^{+} }dz ^{+}f^{bac} A^{-}_b (z )\}$ is the gauge link in
the adjoint representation. This gluon distribution can also be defined in the fundamental
representation~\cite{Mulders:2000sh},
\begin{eqnarray}
 x'G( x' )&=&2\int \frac{d\xi ^{+}}{2\pi \bar P^{-}}%
e^{-ix'\bar P^{-}\xi ^{+}} \langle P|{\rm Tr}\left[F^{-\mu}_a(\xi ^{+})T^a \mathcal{L}
F^{-\mu}_c(0) T^c\mathcal{L}^\dag \right]|P\rangle \ ,\label{g1fund}
\end{eqnarray}
where the gauge link takes form $\mathcal{L}=\mathcal{P}\exp\{-ig\int_{0}^{\xi ^{+} }dz ^{+}T^b
A^{-}_b (z )\}$.
\begin{figure}[hbtp]
\begin{center}
\includegraphics[angle=0,scale=0.75]{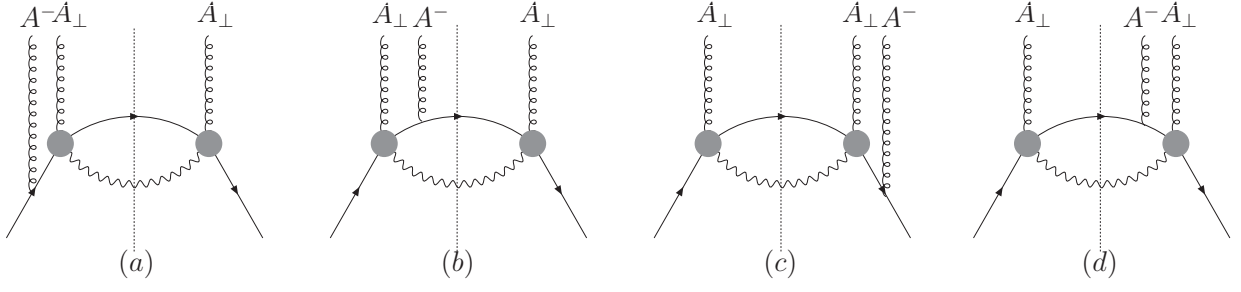}
\caption{The lowest order diagrams contributing to the gauge link of the integrated gluon
distribution. } \label{1}
\end{center}
\end{figure}

The above gauge link is built up by summing longitudinal gluon($A^-$) attachment to all orders.  As
a warm up exercise, we first rederive the gauge link at lowest nontrivial order by computing the
diagrams illustrated in Fig.2. The gluon pole and the color structure associated with the initial
state interaction in the amplitude is given by,
\begin{eqnarray}
 \frac{1}{x'_g+i\epsilon} {\rm Tr}\left [ T^a T^c T^b\right ]
\end{eqnarray}
  which yields the following contribution to the first order expansion of the gauge link,
\begin{eqnarray}
 \langle P|{\rm Tr}\left[F^{-\mu}_a(\xi ^{+})T^a  F^{-\mu}_c(0)
T^c \left ( -ig\int_{-\infty}^{0 }dz^{+}T^b A^{-}_b (z )\right ) \right]|P\rangle
\end{eqnarray}
 And similarly, for the final state interaction in the amplitude, one has,
\begin{eqnarray}
\frac{1}{-x'_g+i\epsilon}{\rm Tr} \left [ T^a T^b T^c\right ] \ \ \Rightarrow \ \ \
 \langle P|{\rm Tr}\left[F^{-\mu}_a(\xi ^{+})T^a
 \left (-ig\int_{0}^{\infty }dz ^{+}T^b A^{-}_b (z )\right ) F^{-\mu}_c(0)
T^c \right]|P\rangle
\end{eqnarray}
 For Fig.2(c), one has,
\begin{eqnarray}
 \frac{1}{x'_g-i\epsilon} {\rm Tr}\left [ T^b T^a T^c \right ]
\ \ \Rightarrow \ \ \
 \langle P|{\rm Tr}\left[ \left (
 ig\int_{-\infty}^{\xi^+}dz^{+}T^b A^{-}_b (z )\right ) F^{-\mu}_a(\xi ^{+})T^a  F^{-\mu}_c(0)
T^c  \right]|P\rangle
\end{eqnarray}
 and Fig.2(d),
\begin{eqnarray}
\frac{1}{-x'_g-i\epsilon}{\rm Tr} \left [ T^a T^b T^c\right ] \ \ \Rightarrow \ \ \
 \langle P|{\rm Tr}\left[F^{-\mu}_a(\xi ^{+})T^a
 \left (ig\int_{\xi^+}^{\infty }dz ^{+}T^b A^{-}_b (z )\right ) F^{-\mu}_c(0)
T^c \right]|P\rangle
\end{eqnarray}
Summing up all terms  gives rise to the first nontrivial order expansion of the gauge link. It is
easy to generalize the above derivation and absorb longitudinal gluon exchange into the gauge link
to all orders. Meanwhile, the gauge link of the integrated quark distribution from the projectile
nucleon is build up by summing longitudinal gluon $A^+$ attachment. One doesn't expect that these
two procedures (summing $A^-$ and $A^+$ gluon exchanges) interfere with each other at leading twist
level.

We now turn to review the conventional twist-3 calculation for the SSA in prompt photon production
process~\cite{Qiu:1991wg,Kouvaris:2006zy}.  In a covariant gauge calculation, as shown in Fig.3 an
additional $A^+$ gluon which carries small transverse momentum $p_\perp$ must be exchanged in order
to generate an imaginary phase necessary for the nonvanishing SSA.  One can isolate  imaginary part
by picking up gluon poles generated via the initial state interactions or the final state
interactions as shown in Fig.3. However, as well known, there is a complete cancelation between the
contributions from the different cut diagrams with an longitudinal gluon attaching to the
unobserved final state produced particle. This can be best seen by explicitly writing down the
gluon pole contribution and  the on shell condition from Fig.3(c),
 \begin{eqnarray}
\frac{1}{(l_q-x_gP-k_\perp)^2+i\epsilon}|_{\text{pole}} \delta(l_q^2)=-i\pi\delta \left
((l_q-x_gP-k_\perp)^2 \right ) \delta(l_q^2)
\end{eqnarray}
and from the left cut diagram Fig.3(d),
\begin{eqnarray}
\frac{1}{l_q^2-i\epsilon}|_{\text{pole}} \delta\left((l_q-x_gP-k_\perp)^2 \right
)=+i\pi\delta(l_q^2)\delta \left ((l_q-x_gP-k_\perp)^2 \right )
\end{eqnarray}
where $l_q=xP+x'\bar P+x_gP+p_\perp-l_\gamma$. Obviously, they cancel out each other as the rest
part of the amplitude squared represented by Fig.3(c) and Fig.3(d) are exactly same. The absence of
the final state interaction contributions to the SSA in the current case is actually the key
observation which leads us to conclude that color entanglement effect also shows up in collinear
twist-3 factorization. We will explain the reasoning in details in the next section.

To isolate the twist-3 effect, one proceeds by expanding the amplitudes squared and the on shell
condition in terms of $p_\perp$,
\begin{eqnarray}
 {\cal H} ( p_\perp)\delta(l_q^2)=
{\cal H} ( p_\perp)\delta(l_q^2)|_{p_\perp=0}
 +\frac{\partial  {\cal H}( p_\perp)\delta(l_q^2)}{\partial p_\perp^\rho}|_{p_\perp=0} \ p_\perp^\rho + ...
\end{eqnarray}
The twist-3 spin dependent part is the term linear in $p_\perp$. In this work, we only take into
account the derivative term contribution, for which case our analysis can be greatly simplified
without losing generality. For the derivative term contribution,  one can simply neglect $p_\perp$
in the hard part ${\cal H}$,
\begin{eqnarray}
  \frac{-l_{\gamma \perp}^\rho}{l_q \cdot P}  {\cal H} ( p_\perp=0)
\left [ \frac{ \partial \delta(l_q^2)}{\partial x}\right ]_{p_\perp=0} \  p_{\perp, \rho}
\end{eqnarray}
One then can carry out integration over $p_\perp$ after converting $A^+$ into the gauge invariant
form $ F^{ \rho  +}$ by partial integration. The corresponding three parton correlation function
can be cast into the form of the ETQS function defined as~\cite{Efremov:1981sh,Qiu:1991pp},
\begin{eqnarray}
T_{F,q}(x_1,x_2)=\int \frac{dy_1^-dy_2^-}{4 \pi} e^{ix_1P^+ y_1^- + i(x_2-x_1)P^+y^-_2} \langle
P,S_\perp | \bar{\psi}_q(0)\gamma^+ g \epsilon^{S_\perp \rho n p} F_ \rho^{ \
+}(y_2^-)\psi_q(y_1^-) | P,S_\perp \rangle
\end{eqnarray}
where we have suppressed gauge links. $S_\perp$ denotes the proton transverse spin vector. Note
that our definition of the ETQS functions differs by a factor $ g$ from the convention used in
Ref.~\cite{Kouvaris:2006zy}.

Making use of the ingredients described above, the calculation is straightforward. The derivative
term contribution to the spin dependent cross section is given
in~\cite{Qiu:1991wg,Kouvaris:2006zy},
\begin{eqnarray}
 \frac{d^3 \Delta \sigma}{d^2 l_{\gamma\perp}dz}&=&
 \frac{\alpha_s \alpha_{em} N_c }{N_c^2-1} \ \frac{(z^2-z)[1+(1-z)^2]}{l_{\gamma\perp}^4}
 \frac{\epsilon^{l_{\gamma} S_\perp  n p}}{ l_{\gamma\perp}^2}
 \sum_q e_q^2 \int^1_{x_{\text{min} }} dx \  x'G(x') \left [-x \frac{d}{dx}
T_{F,q}(x,x)  \right ]
\end{eqnarray}
The same observable has also been computed in a hybrid approach~\cite{Schafer:2014zea}. It has been
found that two approaches doesn't produce the same result. To be more explicit, $ x'G(x')$ in the
above formula is replaced with the combination $ x'G(x')-x'G_4(x')$ in the hybrid approach
calculation in the collinear limit. Numerically, the impact of $G_4$ on the SSA for photon
production is limited as $G_4$ is $1/N_c^2$ suppressed as compared to the normal integrated gluon
distribution $G$~\cite{Schafer:2014zea,Schafer:2014xpa,Zhou:2015ima}. However, a recent
work~\cite{Zhou:2017sdx} shows that the SSA for the forward inclusive jet production in pp/pA
collisions is proportional to the combination $ x'G(x')-N_c^2 x'G_4(x')$, which drastically
deviates from the prediction of the conventional collinear twist-3 calculations. Therefore, from
both theoretical and phenomenological point of view, it is important to pin down the source of the
inconsistency between two approaches. We address this issue in the next section.
\begin{figure}[hbtp]
\begin{center}
\includegraphics[angle=0,scale=0.75]{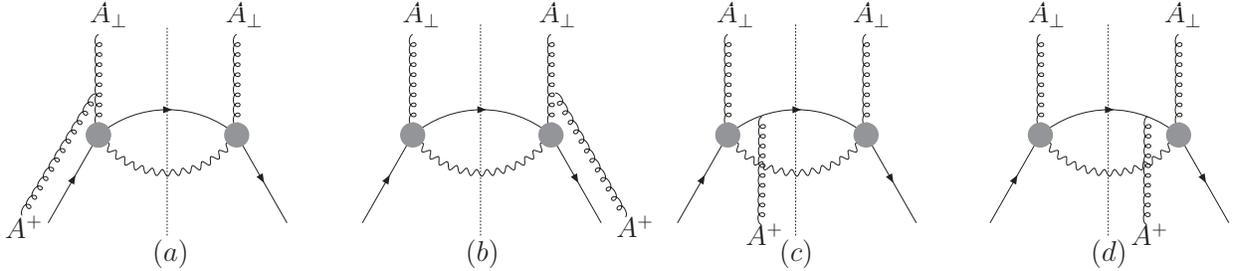}
\caption{Soft gluon pole contributions to the single spin asymmetry for direct photon production in
pp collisions. Final state interaction contributions (c) and (d) to the spin asymmetry cancel out.
Note that $A^+$ gluon carries small transverse momentum in the twist-3 calculation.} \label{1}
\end{center}
\end{figure}

\section{gauge link structure in collinear twist-3  factorization}
The new gluon distribution $G_4$ appears in the hybrid approach calculation results from color
entanglement which arises when considering $A^+$ and $A^- $ gluon exchanges from each side
  simultaneously. The operator definition of the integrated $G_4$ reads~\cite{Schafer:2014xpa},
\begin{eqnarray}
x' G_4(x')=\frac{2}{N_c} \int \frac{d\xi^+ }{2\pi \bar P^-} e^{-ix'\bar P^-\xi^+}
 \langle P| {\rm Tr} \left[ {\cal L}_\xi F^{-\mu}(\xi^+) \right ] {\rm Tr}
\left[ {\cal L}^\dag_0 F^{-\mu}(0) \right ] |P \rangle
\end{eqnarray}
where the gauge link is given by,
\begin{eqnarray}
{\cal L_\xi}&=&{\cal P} {\rm exp} \left [ ig \int_{\xi^+}^{+\infty} dz^+ A^-(z^+,0_\perp) \cdot T
\right ] {\cal P} {\rm exp} \left [ ig \int_{0}^{+\infty} dz_\perp A_\perp(+\infty^+,z_\perp) \cdot
T \right ] \nonumber\\  && \times  \ {\cal P} {\rm exp} \left [ ig \int_{+\infty}^{0} dz_\perp
A_\perp(-\infty^+,z_\perp) \cdot T \right ] {\cal P} {\rm exp} \left [ ig \int_{-\infty}^{\xi^+}
dz^+ A^-(z^+,0_\perp) \cdot T \right ]
\end{eqnarray}
with transverse gauge links~\cite{Ji:2002aa,Belitsky:2002sm} being included. We are now aiming at
recovering the above novel color structure within the pure collinear twist-3 formalism.  Before
making an all order analysis, it is instructive to carry out an explicit calculation at lowest
nontrivial order. To this end, one has to compute diagrams with one $A^-$ gluon attachments on each
side of cut due to ${\rm Tr}[T^a]=0$.  The relevant diagrams are shown in Fig.5.

As we focus on the derivative term contribution, we can neglect transverse momentum $p_\perp$
carried by $A^+$  gluon in the  hard parts except for $p_\perp$ dependence in the on shell
condition $\delta(l_q^2)$. The hard parts computed from different diagrams are then proportional to
the twist-2 spin independent Born diagram contribution, but with different gluon pole and color
structure. Therefore, for the current purpose, it is sufficient to only present the associated
gluon pole and color structure for each diagram shown in Fig.5.

To simplify the calculation of diagrams with three gluon vertex,  it is convenient to organize
$A^+$ and $A^-$ gluon fusion amplitude into two terms,
\begin{eqnarray}
 &&\frac{-i g f^{abc}}{(x_gP+x_g' \bar P)^2 +i\epsilon }
 \left[ g^{\mu \nu}(x_g'\bar P-x_gP)^\rho +g^{\nu \rho}(2x_gP+x'_g \bar P)^\mu
 +g^{\rho \mu}(-2x_g'\bar P-x_g P)^\nu \right ] n^\mu p^\nu
 \nonumber \\ &&
=\frac{-i g f^{abc}\left( x_gP  +x_g' \bar P \right )^\rho}{(x_gP+x_g' \bar P)^2 +i\epsilon }
-\frac{-i g f^{abc} \bar P ^\rho}{x_gP \cdot \bar P +i\epsilon }
\end{eqnarray}
where the first term in the second line doesn't contribute to the final result due to the Ward
identity. For example, the contribution from the first term is canceled  out when summing up
diagrams Fig.5(a), Fig.5(d) and Fig.5(i).   One can use the second term as the effective Feynman
rule for the $A^+$ and $A^-$ gluon fusion vertex  in the following calculation. It is important to
notice that the associated gluon pole $\frac{1} {x_gP \cdot \bar P +i\epsilon }$ corresponds to the
initial state interaction contribution.
\begin{figure}[hbtp]
\begin{center}
\includegraphics[angle=0,scale=0.4]{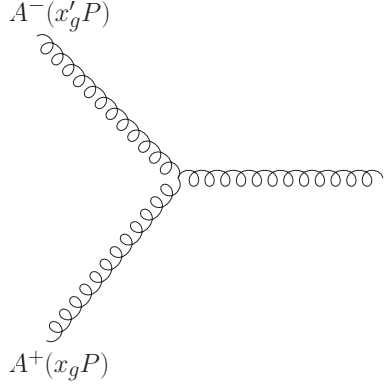}
\caption{ Two longitudinally polarized gluons fusion diagram.} \label{1}
\end{center}
\end{figure}

Using the above trick, it is easy to verify that the summation of the hard parts Fig.5(i), Fig.5(j)
and Fig.5(k) vanishes,
\begin{eqnarray}
{\cal H}_i+{\cal H}_j+{\cal H}_k=0
\end{eqnarray}
And in an exactly analogous fashion, one has,
\begin{eqnarray}
{\cal H}_t+{\cal H}_u+{\cal H}_v=0
\end{eqnarray}
Meanwhile, we also notice that,
\begin{eqnarray}
{\cal H}_h=0 \ , \ \ \ \ \ {\cal H}_s=0
\end{eqnarray}
We continue with the evaluation of Fig.5(a) and Fig.5(b),
\begin{eqnarray}
{\cal H}_a&\propto &\frac{1}{x_g+i \epsilon}\frac{1}{-x'_{g1}-i \epsilon} \frac{1}{x_g'+i\epsilon}
{\rm Tr} \left [ T^a T^b T^c T^f T^e \right ]i f^{def}
\\
{\cal H}_b&\propto&\frac{1}{x_g+i \epsilon}\frac{1}{-x'_{g1}-i \epsilon} \frac{1}{x_g'+i\epsilon}
{\rm Tr} \left [ T^a T^b T^f T^d T^e \right ] if^{cef}
\end{eqnarray}
Summing up them, one obtains,
\begin{eqnarray}
{\cal H}_{a+b}&\propto&\frac{1}{x_g+i \epsilon}\frac{1}{-x'_{g1}-i \epsilon}
\frac{1}{x_g'+i\epsilon} \left\{ C_F{\rm Tr} \left [ T^a T^b T^c T^d \right ]-{\rm Tr} \left [ T^a
T^b T^e T^c T^d T^e \right ] \right \}
\end{eqnarray}
which leads to the following gauge link structure,
\begin{eqnarray}
{\cal H}_{a+b}&\Rightarrow&C_F \langle P|{\rm Tr}\left[F^{-\mu}_a(\xi ^{+})T^a  \left (
ig\int^{\infty}_{\xi^+ }dz^{+}T^b A^{-}_b (z )\right )
 F^{-\mu}_c(0)T^c \left ( -ig\int_{-\infty}^{0 }dz^{+}T^d A^{-}_d (z )\right )
 \right]|P\rangle \nonumber  \\
 &&-\langle P|{\rm Tr}\left[F^{-\mu}_a(\xi ^{+})T^a  \left (
ig\int^{\infty}_{\xi^+ }dz^{+}T^b A^{-}_b (z )\right ) T^e
 F^{-\mu}_c(0)T^c \left ( -ig\int_{-\infty}^{0 }dz^{+}T^d A^{-}_d (z )\right )
 T^e \right]|P\rangle
\end{eqnarray}
The gluon poles and color structures associated with  Fig.5(c) to Fig.5(g) are listed in the
following,
\begin{eqnarray}
{\cal H}_c&\propto &\frac{1}{x_g+i \epsilon}\frac{1}{-x'_{g1}-i \epsilon} \frac{1}{-x_g'+i\epsilon}
{\rm Tr} \left [ T^a T^b T^d T^f T^e \right ] if^{cef}
\\
{\cal H}_d&\propto&\frac{1}{x_g+i \epsilon}\frac{1}{-x'_{g1}-i \epsilon} \frac{2 l_q \cdot \bar
P}{(l_q-x_g'\bar P-x_g P)^2 +i\epsilon} {\rm Tr} \left [ T^a T^b T^f T^c T^e \right ]if^{def}
\\
{\cal H}_e&\propto &\frac{-1}{-x_g+i \epsilon}\frac{1}{-x'_{g1}-i \epsilon}
\frac{1}{x_g'+i\epsilon} {\rm Tr} \left [ T^a T^b T^e T^c T^d T^e\right ]
\\
{\cal H}_f&\propto&\frac{-1}{-x_g+i \epsilon}\frac{1}{-x'_{g1}-i \epsilon} \frac{2 l_q \cdot \bar
P}{(l_q-x_g'\bar P-x_g P)^2 +i\epsilon} {\rm Tr} \left [ T^a T^b T^e T^d T^c T^e\right ]
\\
{\cal H}_g&\propto & \frac{-1}{-x'_g+i \epsilon}\frac{1}{-x'_{g1}-i \epsilon} \frac{2 l_q \cdot
 P}{(l_q-x_g'\bar P-x_g P)^2 +i\epsilon} {\rm Tr} \left [ T^a T^b T^d T^e T^c T^e \right ]
\end{eqnarray}
We further decompose the hard parts  ${\cal H}_d$ and ${\cal H}_f$ into two terms ${\cal H}_d={\cal
H}_{d1}+{\cal H}_{d2}$ and ${\cal H}_f={\cal H}_{f1}+{\cal H}_{f2}$, where ${\cal H}_{d1}, \ {\cal
H}_{f1}$ denote contributions by picking up the gluon poles: $\frac{1}{x_g+i \epsilon}$ and
$\frac{1}{-x_g+i \epsilon}$, while the imaginary phase of  ${\cal H}_{d2}, \ {\cal H}_{f2}$ is only
generated from the gluon pole $\frac{1}{(l_q-x_g'\bar P-x_g P)^2 +i\epsilon}$.  When we isolate the
imaginary phase from the pole $\frac{1}{(l_q-x_g'\bar P-x_g P)^2 +i\epsilon}$, an internal quark
propagator is effectively put on shell.  The Ward identity then implies that,
\begin{eqnarray}
{\cal H}_{d2}+{\cal H}_{f2}+{\cal H}_g=0
\end{eqnarray}
This relation also can be readily verified by  explicit calculation. The hard parts ${\cal H}_{f1}$
and ${\cal H}_e$ represent the final state interaction contributions, which are canceled out by the
corresponding left cut diagrams, as explained in the previous section. We are now only left with
\begin{eqnarray}
{\cal H}_{c+d1} \propto \frac{1}{x_g+i \epsilon}\frac{1}{-x'_{g1}-i \epsilon}
\frac{1}{-x_g'+i\epsilon} \left\{ C_F{\rm Tr} \left [ T^a T^b T^d T^c \right ]-{\rm Tr} \left [ T^a
T^b T^e T^d T^c T^e \right ] \right \}
\end{eqnarray}
which results in,
\begin{eqnarray}
{\cal H}_{c+d1}&\Rightarrow&C_F \langle P|{\rm Tr}\left[F^{-\mu}_a(\xi ^{+})T^a  \left (
ig\int^{\infty}_{\xi^+ }dz^{+}T^b A^{-}_b (z )\right )
 \left ( -ig\int^{\infty}_{0 }dz^{+}T^d A^{-}_d (z )\right )F^{-\mu}_c(0)T^c
 \right]|P\rangle \nonumber  \\
 &&-\langle P|{\rm Tr}\left[F^{-\mu}_a(\xi ^{+})T^a  \left (
ig\int^{\infty}_{\xi^+ }dz^{+}T^b A^{-}_b (z )\right ) T^e
 \left ( -ig\int^{\infty}_{0 }dz^{+}T^d A^{-}_d (z )\right ) F^{-\mu}_c(0)T^c
 T^e \right]|P\rangle
\end{eqnarray}

The similar analysis applies to the rest of diagrams.  One finds that
\begin{eqnarray}
{\cal H}_{o2}+{\cal H}_{q2}+{\cal H}_r=0
\end{eqnarray}
and the contributions from ${\cal H}_{q1}$ and ${\cal H}_p$ drop out once combining with the
corresponding left cut diagrams.
\begin{figure}[hbtp]
\begin{center}
\includegraphics[angle=0,scale=0.8]{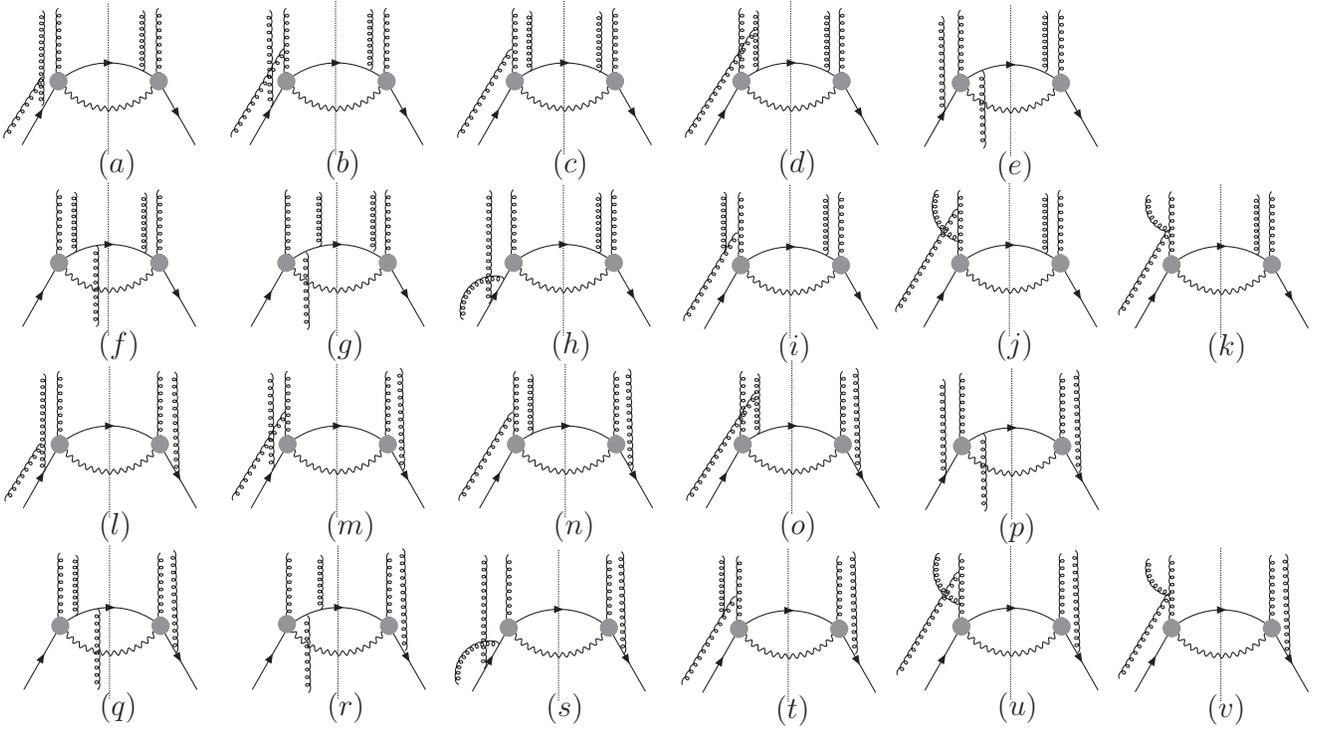}
\caption{The lowest nontrivial order diagrams giving rise to color entanglement effect. The mirror
diagrams are not shown here. Gluons directly attaching to grey circles are transversely polarized.
Other gluons from the top part of diagrams are longitudinally polarized $A^-$, while  $A^+$ gluon
attaches hard parts from the bottom part.} \label{5}
\end{center}
\end{figure}
One eventually has,
\begin{eqnarray}
{\cal H}_{l+m}&\Rightarrow&C_F \langle P|{\rm Tr}\left[ \left ( ig\int_{-\infty}^{\xi^+ }dz^{+}T^b
A^{-}_b (z )\right ) F^{-\mu}_a(\xi ^{+})T^a
 F^{-\mu}_c(0)T^c \left ( -ig\int_{-\infty}^{0 }dz^{+}T^d A^{-}_d (z )\right )
 \right]|P\rangle \nonumber  \\
 &&-\langle P|{\rm Tr}\left[ \left (
ig\int_{-\infty}^{\xi^+ }dz^{+}T^b A^{-}_b (z )\right ) F^{-\mu}_a(\xi ^{+})T^a  T^e
 F^{-\mu}_c(0)T^c \left ( -ig\int_{-\infty}^{0 }dz^{+}T^d A^{-}_d (z )\right )
 T^e \right]|P\rangle
\end{eqnarray}
and
\begin{eqnarray}
{\cal H}_{n+o1}&\Rightarrow&C_F \langle P|{\rm Tr}\left[\left ( ig\int_{-\infty}^{\xi^+ }dz^{+}T^b
A^{-}_b (z )\right ) F^{-\mu}_a(\xi ^{+})T^a
 \left ( -ig\int^{\infty}_{0 }dz^{+}T^d A^{-}_d (z )\right )F^{-\mu}_c(0)T^c
 \right]|P\rangle \nonumber  \\
 &&-\langle P|{\rm Tr}\left[\left ( ig\int_{-\infty}^{\xi^+ }dz^{+}T^b
A^{-}_b (z )\right ) F^{-\mu}_a(\xi ^{+})T^a   T^e
 \left ( -ig\int^{\infty}_{0 }dz^{+}T^d A^{-}_d (z )\right ) F^{-\mu}_c(0)T^c
 T^e \right]|P\rangle
\end{eqnarray}

Collecting all pieces ${\cal H}_{a+b},{\cal H}_{c+d1}, {\cal H}_{l+m}, {\cal H}_{n+o1}$ together,
it is easy to see that the summation of them gives rise to the lowest nontrivial order expansion of
the following gauge link structure,
\begin{eqnarray}
 &&C_F \langle P|{\rm Tr}\left[F^{-\mu}_a(\xi ^{+})T^a \mathcal{L}(\xi^+,0) F^{-\mu}_c(0)
T^c\mathcal{L}^\dag(\xi^+,0) \right]|P\rangle \nonumber
\\ &&-\langle P|{\rm Tr}\left[\mathcal{L}^\dag(-\infty,\xi^+)
F^{-\mu}_a(\xi ^{+})T^a \mathcal{L}^\dag(\xi^+,\infty) T^e \mathcal{L}(\infty,0)F^{-\mu}_c(0)
T^c\mathcal{L}(0,-\infty) T^e \right]|P\rangle \label{37}
\end{eqnarray}
where the term in the second line has quite peculiar color structure. Note that the final state
interaction ${\cal H}_{q1}$, ${\cal H}_p$, ${\cal H}_{f1}$ and ${\cal H}_e$ contribute to the
twist-2 unpolarized cross section.  The principal value part of the gluon poles $\frac{1}{-x_g+i
\epsilon}$ and $\frac{1}{x_g+i \epsilon}$  is canceled out when adding up these contributions with
the second term in the above formula. Correspondingly, $A^+$ gluon attachment is decoupled from the
hard part and can be absorbed into the gauge link of the collinear quark distribution. This leads
us to conclude that the collinear twist-2 factorization is not affected by color entanglement
effect at the order under consideration. Meanwhile, it  becomes clear that the emergence of color
entanglement effect in the collinear twist-3 factorization essentially can be attributed to the
fact that the final state interaction doesn't contribution to the SSA. With this observation, one
can generalize the above analysis to all orders. Considering a diagram with arbitrary number of
$A^-$ gluon attachments and one $A^+$ attachment from the opposite side, $A^+$ gluon can be first
decoupled from the hard part by invoking the Ward identity argument once we sum over all possible
$A^+$ insertion points. After having done this, multiple $A^-$ exchange can be incorporated into
the normal gauge link as shown in the first line of Eq.~\ref{37}. However, one has to keep in mind
that only initial state interaction contributes to the SSA in the current case. The gauge link
structure associated with final state interaction given in the seconde line of Eq.~\ref{37} has to
be subtracted. The overall color structure associated with initial state interaction is thus given
by Eq.~\ref{37}. Note that the similar argument also applies to the SSA in other channels, for
instance, inclusive hadron production in pp collisions, because initial state interaction and final
state interaction don't generate contributions with equal weight.

Eq.~\ref{37} can be further cast into the following form using the Fierz identity,
\begin{eqnarray}
 &&\frac{N_c}{2} \langle P|{\rm Tr}\left[F^{-\mu}_a(\xi ^{+})T^a \mathcal{L}(\xi^+,0) F^{-\mu}_c(0)
T^c\mathcal{L}^\dag(\xi^+,0) \right]|P\rangle \nonumber \\ &&-\frac{1}{2} \langle P|{\rm
Tr}\left[\mathcal{L}^\dag(-\infty,\xi^+) F^{-\mu}_a(\xi ^{+})T^a
\mathcal{L}^\dag(\xi^+,\infty)\right ] {\rm Tr} \left [ \mathcal{L}(\infty,0)F^{-\mu}_c(0)
T^c\mathcal{L}(0,-\infty)  \right]|P\rangle
\end{eqnarray}
One immediately recognizes this operator structure as the one that gives rise to the combination
$x' G(x')-x' G_4(x')$. The corresponding spin dependent cross section reads,
\begin{eqnarray}
 \frac{d^3 \Delta \sigma}{d^2 l_{\gamma\perp}dz}=
 \frac{\alpha_s \alpha_{em} N_c }{N_c^2-1} \ \frac{(z^2-z)[1+(1-z)^2]}{l_{\gamma\perp}^4}
 \frac{\epsilon^{l_{\gamma} S_\perp  n p}}{ l_{\gamma\perp}^2}
 \sum_q e_q^2 \int^1_{x_{\text{min} }} \! dx \left [ x' G(x')-x' G_4(x') \right ] \left [-x \! \frac{d}{dx}
T_{F,q}(x,x)  \right ]
\end{eqnarray}
which is in full agreement with that obtained by extrapolating the hybrid approach result to the
collinear limit~\cite{Schafer:2014zea}. As expected, the pure collinear twist-3
 formalism and the hybrid approach are
consistent with each other in the collinear limit after properly taking into account color
entanglement effect in collinear twist-3 formalism.

\section{Summary}
In summary, we have shown that color entanglement effect plays a role in contributing to T-odd
observables within the  genuine collinear twist-3 factorization framework.  As an example, we
compute the SSA for direct photon production in pp collisions using the collinear twist-3 approach.
Our calculation differs from the conventional collinear twist-3 treatment by deriving the gauge
link structure on unpolarized target side explicitly.  We found that the existence of the
longitudinal gluon $A^+$ attachment from the polarized projectile leads to a peculiar gauge link
structure when summing up $A^-$ gluon exchanges, which can be summarized into the novel gluon
distribution $G_4$. As expected,  pure collinear twist-3 formalism and the hybrid approach do yield
the same result in the overlap region where they both apply.

In the present work, we only focus on the derivative term contribution. But we anticipate that the
non-derivative term contribution and the soft fermion pole contribution~\cite{Kanazawa:2012kt} are
also affected by color entanglement effect. Moreover,  it has been found in the hybrid approach
calculations~\cite{Schafer:2014xpa,Zhou:2015ima,Zhou:2017sdx} that color entanglement effect also
contributes to the SSAs in other processes in pp/pA collisions. It is of great interest to confirm
these results within  pure collinear twist-3 approach, particularly in view of that color
entanglement effect plays a crucial role in solving the sign mismatch problem~\cite{Zhou:2017sdx}.
In a word, all previous collinear twist-3 calculations for T-odd observables(including these
related to the Boer-Mulders effect~\cite{Boer:1997nt,Zhou:2008fb,Zhou:2009rp,Kanazawa:2014nea})
should be thoroughly re-examined by analyzing the gauge link structure of the collinear twist-2
parton distributions on target side.

There are also some other theoretical issues remain to be addressed in  future study. First of all,
one has to investigate if all pure gauge  gluons $A^+$ attachments other than the one carried small
transverse momentum can be decoupled from hard parts and absorbed into the gauge link in the ETQS
function. If true,  relevant cross sections are factorizable in  collinear twist-3 factorization.
Otherwise, collinear twist-3 factorization breaks down for T-odd observables in pp collisions.
Second, it is worthwhile to make effort to study the properties of the gluon distribution $G_4$ at
moderate or large $x$, such as its scale evolution, behavior in some models. Finally, though we
tend to believe that the conventional collinear twist-3 formulation of T-even
observables~\cite{Zhou:2009jm,Liang:2012rb,Metz:2012fq,Hatta:2013wsa,Koike:2016ura} and
fragmentation
effects~\cite{Yuan:2009dw,Kang:2010zzb,Zhou:2011ba,Metz:2012ct,Kanazawa:2014dca,Hatta:2016khv,Gamberg:2017gle,Metz:2016swz}
are not affected by color entanglement effect, it would be nice to verify this point by explicit
calculations.

\

\noindent {\it \bf Acknowledgments:} This work has been supported by  the National Science
Foundation of China under Grant No. 11675093, and by the Thousand Talents Plan for Young
Professionals.

\end {document}